\documentclass{svproc}
\usepackage{url}
\usepackage{booktabs} 
\usepackage{subfig}
\usepackage{graphicx}
\usepackage{booktabs,caption}
\usepackage[flushleft]{threeparttable}
\usepackage{svg}
\usepackage{appendix}
\usepackage{rotating}

\begin{document}
\mainmatter              
\title{Combining STPA and BDD for Safety Analysis and Verification in Agile Development: A Controlled Experiment}
\titlerunning{Combining STPA and BDD in Agile Development}  
%
\author{Yang Wang \and Stefan Wagner}
\authorrunning{Yang Wang and Stefan Wagner} 
%
\institute{University of Stuttgart, Germany \\
\email{\{yang.wang, stefan.wagner\}@informatik.uni-stuttgart.de}}

\maketitle              

\begin{abstract}

\emph{Context:} Agile development is in widespread use, even in safety-critical domains.      
\emph{Motivation:} However, there is a lack of an appropriate safety analysis and verification method in agile development.      
\emph{Objective:} In this paper, we investigate the use of Behavior Driven Development (BDD) instead of standard User Acceptance Testing (UAT) for safety verification with System-Theoretic Process Analysis (STPA) for safety analysis in agile development.  
\emph{Method:} We evaluate the effect of this combination in a controlled experiment with 44 students in terms of productivity, test thoroughness, fault detection effectiveness and communication effectiveness.    
\emph{Results:} The results show that BDD is more effective for safety verification regarding the impact on communication effectiveness than standard UAT, whereas productivity, test thoroughness and fault detection effectiveness show no statistically significant difference in our controlled experiment.
\emph{Conclusion:} The combination of BDD and STPA seems promising with an enhancement on communication, but its impact needs more research.

\end{abstract}
\vspace{-0.2cm}
\section{Introduction}
\vspace{-0.2cm}
Agile practices have been widely used in software industries to develop systems on time and within budget with improved software quality and customer satisfaction \cite{dybaa2008empirical}. The success of agile development has led to a proposed expansion to include safety-critical systems (SCS) \cite{cleland2017case}. However, to develop SCS in an agile way, a significant challenge exists in the execution of safety analysis and verification \cite{arthur2017applying}. The traditional safety analysis and verification techniques, such as failure mode effect analysis (FMEA) and fault tree analysis (FTA) are difficult to apply within agile development. They need a detailed and stable architecture \cite{fleming2015safety}. \par
In 2016, we proposed to use System-Theoretic Process Analysis (STPA) \cite{leveson2011engineering} in agile development for SCS \cite{wang2016toward}. First, STPA can be started without a detailed and stable architecture. It can guide the design. In agile development, a safety analyst starts with performing STPA on a high-level architecture and derives the relevant safety requirements for further design. Second, Leveson developed STPA based on the systems theoretic accident modeling and processes (STAMP) causality model, which considers safety problems based on system theory rather than reliability theory. In today's complex cyber-physical systems, accidents are rarely caused by single component or function failures but rather by component interactions, cognitively complex human decision-making errors and social, organizational, and management factors \cite{leveson2011engineering}. System theory can address this. \par
The safety requirements derived from STPA need verification. However, there is no congruent safety verification in agile development. Most agile practitioners mix unit test, integration test, field test and user acceptance testing (UAT) to verify safety requirements \cite{cleland2017case}. In 2016, we proposed using model checking combined with STPA in a Scrum development process \cite{wang2016towards}. However, using model checking, a suitable model is necessary but usually not available in agile development. In addition, the formal specification increases the difficulties of communication, which should not be neglected when developing SCS \cite{martins2017requirements}. BDD, as an agile technique, is an evolution of test driven development (TDD) and acceptance test driven development (ATDD). The developers repeat coding cycles interleaved with testing. TDD starts with writing a unit test, while ATDD focuses on capturing user stories by implementing automated tests. BDD relies on testing system behavior in scenarios by implementing a template: Given[Context], When[Event], Then[Outcome] \cite{north2012jbehave}. The context describes pre-conditions or system states, the event describes a trigger event, and the outcome is an expected or unexpected system behavior. It could go further into low-level BDD\footnote{\scriptsize{Low-level BDD is possible to define low-level specifications and interwined with TDD \cite{smart2015bdd}.}}. Yet, it has not been used to verify safety requirements. Leveson said \cite{leveson2011engineering}: ``\emph{Accidents are the result of a complex process that results in system behavior violating the safety constraints.}" Hence, in agile development, we need safety verification to: (1) be able to guide design at an early stage, (2) strengthen communication and (3) focus on verifying system behavior. Thus, we believe that BDD might be suitable for safety verification with STPA for safety analysis in agile development.  \newline \newline 
\textbf{Contributions} \newline
We propose a possible way to use BDD with STPA for safety verification in agile development. We investigate its effects regarding productivity, test thoroughness, fault detection effectiveness and communication effectiveness by conducting a controlled experiment with the limitation that we execute BDD only in a test-last way. The results show that BDD is able to verify safety requirements based on system theory, and is more effective than UAT regarding communication for safety verification.  \newline
\vspace{-0.2cm}
\section{Related Work}
\vspace{-0.2cm}
Modern agile development processes for developing safety-critical systems (SCS) advocate a hybrid mode through alignment with standards like IEC 61508, ISO 26262 and DO-178. There have been many considerable successes \cite{vuori2011agile} \cite{staalhane2012application} \cite{ge2010iterative}. However, a lack of integrated safety analysis and verification to face the changing architecture through each short iteration is a challenge for using such standards. In 2016, we proposed to use STPA in a Scrum development process \cite{wang2016toward}. It showed a good capability to ensure agility and safety in a student project \cite{wang2017exploratory}. However, we verified the safety requirements only at the end of each sprint by executing UAT together with TDD in development. A lack of integrated safety verification causes some challenges, such as poor verification and communication.  
The previous research regarding safety verification in agile development suggested using formal methods \cite{eleftherakis2003agile} \cite{ghezzi2013requirements}. However, they need models and make intuitive communication harder \cite{wang2016towards}. In addition, they have not considered specific safety analysis techniques. \par
Hence, we propose using BDD to verify safety requirements. BDD is specifically for concentrating on behavior testing \cite{wynne2012cucumber}. It allows automated testing against multiple artifacts throughout the iterative development process \cite{silva2017behavior}. Moreover, it bridges the gap between natural language-based business rules and code language \cite{hummel2013role}. Okubo et al. \cite{okubo2014security} mentioned the possibilities of using BDD for security and privacy acceptance criteria. They define the acceptance criteria by creating a threat and countermeasure graph to write attack scenarios. They verify the satisfication of security requirements by testing the countermeasures, to see whether they can make the attack scenarios or unsecure scenarios fail. Lai, Leu, and Chu \cite{lai2014combining} combined BDD with iterative and incremental development specifically for security requirements evaluation. They defined the behavioral scenarios by using use case diagram and misuse case diagram. STPA encompasses determining safe or unsafe scenarios. We aim to use BDD verifying these scenarios. \par
To investigate the effect of using BDD for safety verification, we design a controlled experiment referring to a set of TDD experiments. Erdogmus, Morisio and Torchiano \cite{sdf} conducted an experiment with undergraduate students regarding programmer productivity and external quality in an incremental development process. For safety verification in agile development, a high productivity of safety test cases promotes high safety. Madeyski \cite{madeyski2010impact} conducted an experiment comparing ``test-first" and ``test-last" programming practices with regard to test thoroughness and fault detection effectiveness of unit tests. BDD for safety verification covers also low-level tests. Thus, we decided to investigate productivity, test thoroughness and fault detection capability in this experiment. \cite{fucci2013replicated} \cite{fucci2016dissection} \cite{panvcur2011impact} \cite{george2004structured} \cite{siniaalto2007comparative} provided evidence of using these three measures. In addition, George and Williams \cite{george2004structured} focused on the understandability of TDD from the developer's viewpoint. Using BDD for safety verification, we notice the importance of communication between developers and business analysts. We investigate understandability in the measure of communication effectiveness.

\section{STPA Integrated BDD for Safety Analysis and Verification (STPA-BDD)}
\vspace{-0.2cm}

\begin{figure}[!hbpt]
\center
\includegraphics[width=0.8\textwidth]{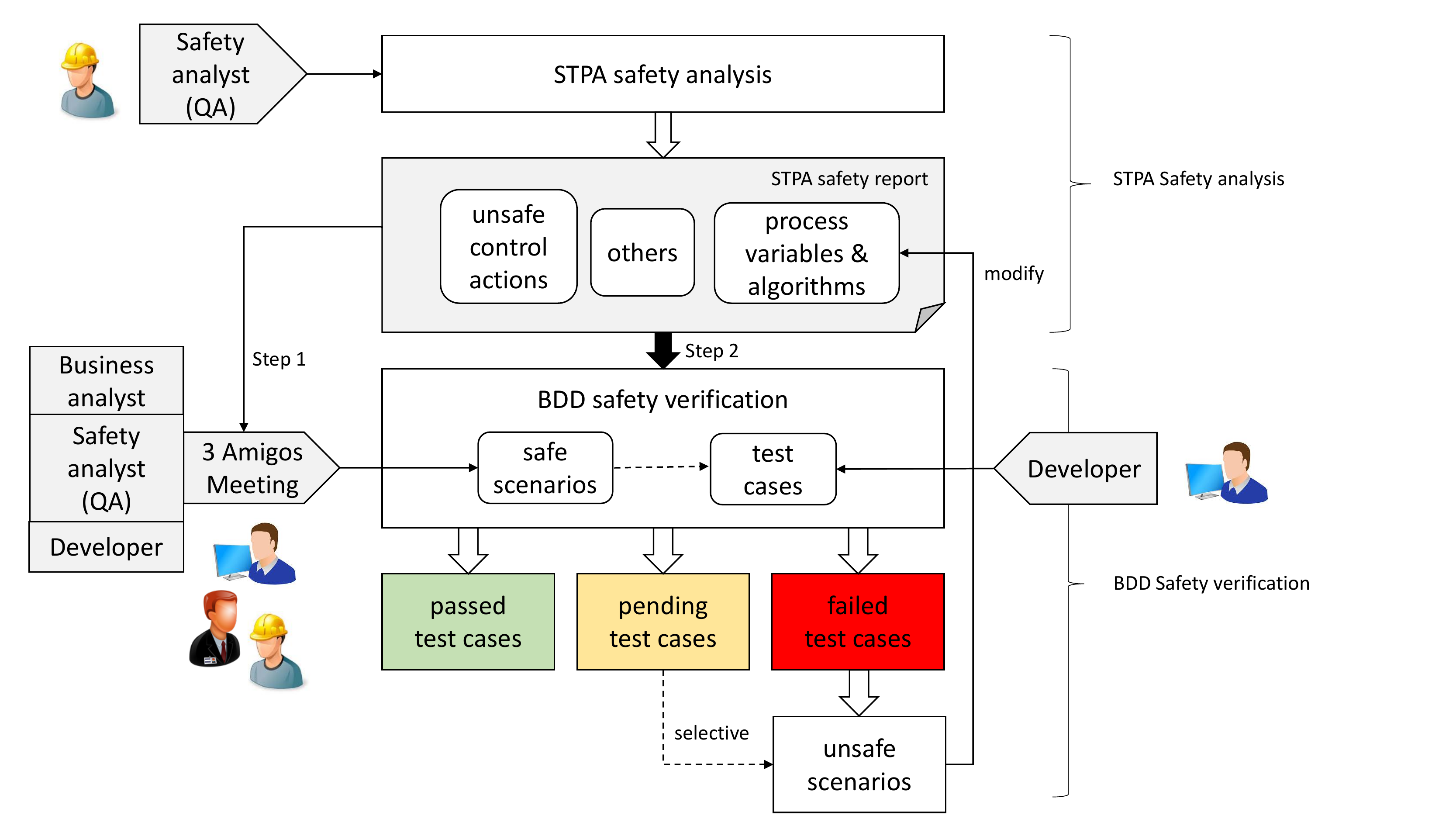}
\caption{STPA-BDD Concept}
\end{figure}

In this article, we propose STPA-BDD. We mainly focus on safety verification.  
As we can see in Fig. 1, we have two main parts: STPA safety analysis and BDD safety verification. A safety analyst\footnote{\scriptsize{Since we focus on safety in our research, we assign a safety analyst as the QA role in our context.}} (QA) starts performing STPA safety analysis with a sufficient amount of code\footnote{\scriptsize{More descriptions of STPA for safety analysis are given in \cite{wang2016towards} concerning an example of using STPA in an airbag system and \cite{wang2017exploratory} concerning the use of STPA in a Scrum development process.}}. STPA is executed by firstly identifying potentially hazardous control actions, and secondly determining how unsafe control actions (UCAs) could occur. STPA derives the safety requirements, which constraint the UCAs, as well as system behaviors. Additionally, it explores the causal factors in scenarios for each UCA. The output from the safety analyst (QA) is an STPA safety report with system description, control structure, accidents,  hazards, UCAs, corresponding safety requirements, process variables and algorithms. \par 
In BDD safety verification, to generate and test scenarios, the UCAs (in STPA step 1), process variables and algorithms (in STPA step 2) from the STPA safety report are needed. We write other data into ``others". 
BDD safety verification has two steps: In step 1, the business analyst, the safety analyst (QA) and the developer establish a ``3 Amigos Meeting" to generate test scenarios. In a BDD test scenario\footnote{\scriptsize{We illustrate a BDD test scenario using only three basic steps ``Given" ``When" ``Then". More annotations, such as ``And", can also be added.}}, we write the possible trigger event for the UCA in \textbf{When [Event]}. The other process variables and algorithms are arranged in \textbf{Given [Context]}. \textbf{Then [Outcome]} presents the expected behavior - a safe control action. In Fig. 2 (a), we present an example. The safety analyst (QA) has provided a UCA as \emph{During auto-parking, the autonomous vehicle does not stop immediately when there is an obstacle upfront.} One of the process variables with relevant algorithms detects the forward distance by using an ultrasonic sensor. The developer considers a possible trigger as the ultrasonic sensor provides the wrong feedback. Thus, a BDD test scenario should test if \emph{the ultrasonic sensor provides the feedback that the forward distance $<=$ threshold (means there is an obstacle upfront)} and whether the vehicle stops. They write this after \textbf{When}. The context could be \emph{the autonomous vehicle is auto-parking.} We write them after \textbf{Given}. \textbf{Then} constraints the safe control action as \emph{the autonomous vehicle stops immediately}. More possible triggers are expected to be generated after \textbf{When} to test them. 
In step 2, after the three amigos discuss and determine the test scenarios, the developer starts generating them into test cases, as shown in Fig. 2 (b). BDD test cases use annotations such as \textbf{@Given}, \textbf{@When}, and \textbf{@Then} to connect the aforementioned test scenarios with real code. The developer produces code to fulfill each annotation. We can identify unsafe scenarios when the test cases fail. We correct the trigger event to pass the test cases to satisfy the safety requirement. \par      
\vspace{-0.2cm}
\begin{figure}[!hbt] 
\centering
\subfloat[Test scenario example]{%
\includegraphics[width=.48\textwidth]{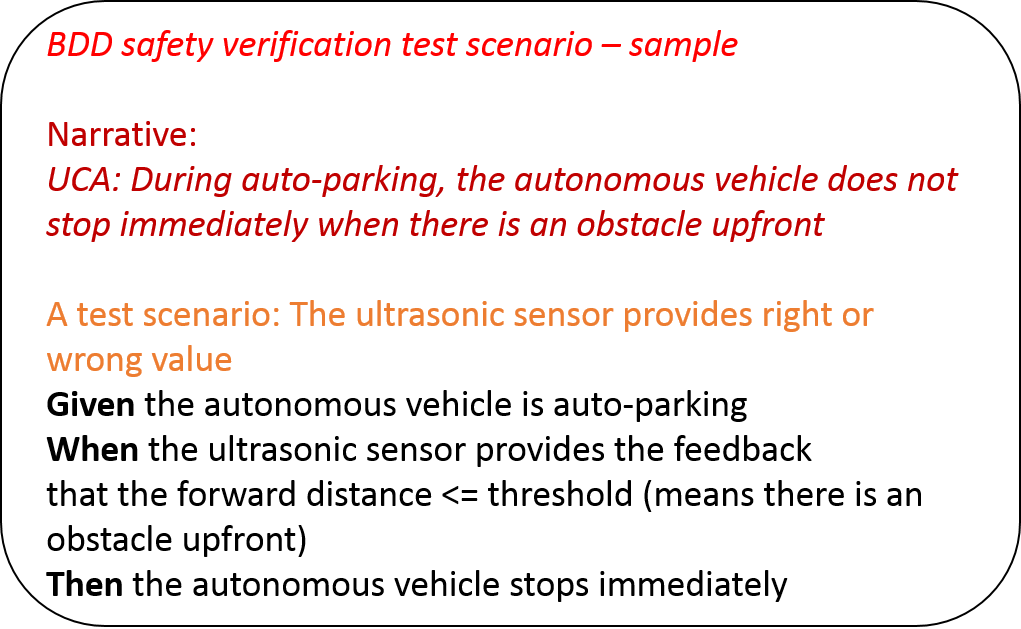}} \vspace{0.2cm}
\subfloat[Test case example]{%
\includegraphics[width=.48\textwidth]{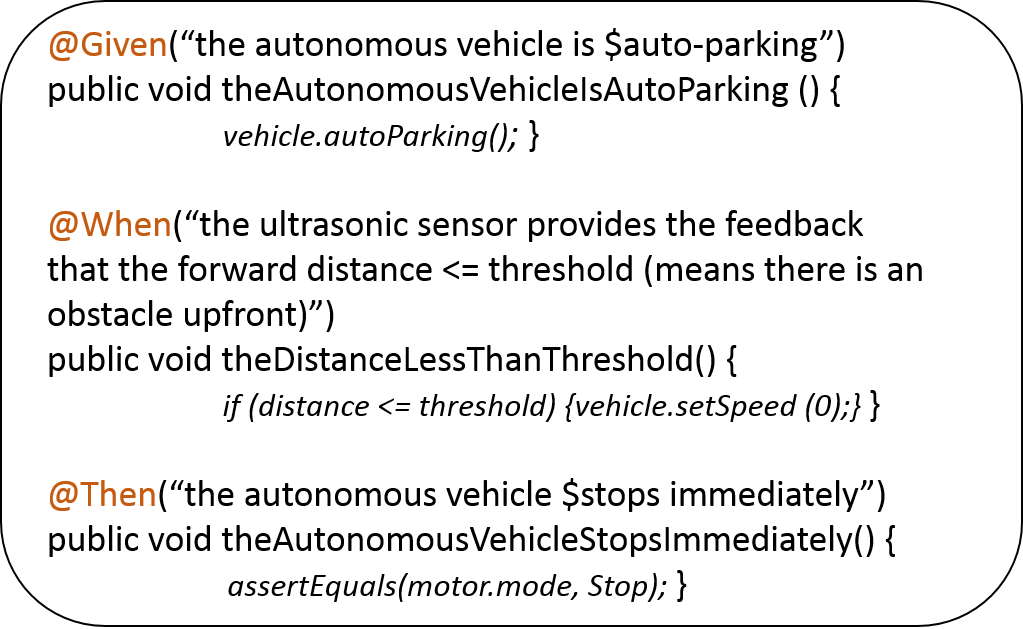}}\\
\caption{BDD safety verification example} 
\end{figure}

\vspace{-0.2cm}
\section[Experiment Design]{Experiment Design\footnote{\scriptsize{We follow the guideline by Wohlin et al. \cite{wohlin2012experimentation}.}}}
\vspace{-0.2cm}

\subsection{Goal}
\vspace{-0.2cm}
\textbf{Analyze} BDD\footnote{\scriptsize{We have a limitation in our experiment that we execute BDD only in a test-last way. More discussion of this issue can be found in Section 6.2}} and UAT\footnote{\scriptsize{To execute a standard UAT, we mainly refer to \cite{crispin2009agile} with fictional business analysts.}} for safety verification. \newline 
\textbf{For the purpose of} comparing their effect. \newline
\textbf{With respect to} 
\emph{productivity} by measuring the number of implemented (tested) user stories per minute; 
\emph{test thoroughness} by measuring line coverage;
\emph{fault detection effectiveness} by measuring a mutation score indicator; 
\emph{communication effectiveness} by conducting a post-questionnaire.\newline
\textbf{From the point of view} of the developers and business analysts.\newline
\textbf{In the context of} B.Sc students majoring in software engineering or other related majors executing acceptance testing. 

\vspace{-0.2cm}
\subsection{Context}
\vspace{-0.2cm}
\emph{Participants}: The experiment ran off-line in a laboratory setting in an ``Introduction to Software Engineering" course at the University of Stuttgart. Since the course includes teaching BDD and UAT technology, the students are suitable subjects for our experiment. We arrange them based on Java programming experiences (not randomly).
According to a pre-questionnaire\footnotemark[14], 88.6\% of the students are majoring in software engineering. We conclude from Table 1 that they have attended relevant lectures and handled practical tasks relating to Java programming, acceptance testing, SCS (with a median value $>$= 3 on a scale from 1 to 5). The agile techniques show less competency (with a median value of 2 on a scale from 1 to 5). We provide a 1-to-1 training, which lasts 44 hours overall, to reduce the weaknesses.  \newline   
\emph{Development environment}: We use a simplified Java code with mutants from a Lego Mindstorms based Autonomous Parking System (APS) and Crossroad Stop and Go System (CSGS). These two systems are comparable by lines of code and number of functional modules\footnotemark[14]. To ease writing test cases, we use a lejo TDD wrapper, Testable Lejos\footnote{\scriptsize{http://testablelejos.sourceforge.net/}} to remove deep dependencies to the embedded environment. 
The BDD groups (Group A1 and Group A2) operate in an Eclipse IDE together with a JBehave plug-in (based on JUnit)\footnote{\scriptsize{http://jbehave.org/eclipse-integration.html}}. We use Eclipse log files and JUnit test reports for calculating the number of implemented (tested) user stories. Finally, we use PIT Mutation Testing\footnote{\scriptsize{http://pitest.org/}} to assess line coverage and a mutation score indicator. The UAT groups (Group B1 and Group B2) write the test cases in Microsoft Word.  \newline 
\vspace{-0.2cm}
\begin{threeparttable}[!hpbt]
\centering
\scriptsize
\caption{Medians of the student's background}
\scriptsize
\begin{tabular}{p{4cm}p{2cm}p{2cm}p{2cm}p{2cm}}
\toprule
  Area & Group A1 & Group A2 & Group B1 & Group B2  \\ \hline 
  Java programming & 3 & 3 & 3 & 3  \\ 
  Acceptance testing &  4   & 5 & 3 & 3  \\ 
  Safety-critical systems & 3 & 4 & 4 & 4  \\ 
  Agile techniques & 3    & 3 & 3 & 2  \\ 
\toprule
\end{tabular}
\begin{tablenotes}
      \tiny
      \item Note: The values range from ``1" (little experience) to ``5" (experienced). Group A1 and Group A2 use BDD, while Group B1 and Group B2 use UAT. 
    \end{tablenotes}
\end{threeparttable} 
\vspace{-0.2cm}         

\vspace{-0.2cm}
\subsection{Hypotheses}
\vspace{-0.2cm}
We formulate the null hypotheses as:  \newline
\textbf{$H_0$ $_{PROD}$}: There is no difference in productivity between BDD and UAT. \newline
\textbf{$H_0$ $_{THOR}$}: There is no difference in test thoroughness between BDD and UAT.\newline
\textbf{$H_0$ $_{FAUL}$}: There is no difference in fault detection effectiveness between BDD and UAT.\newline
\textbf{$H_0$ $_{COME}$}: There is no difference in communication effectiveness between BDD and UAT.\newline
The alternative hypotheses are: \newline
\textbf{$H_1$ $_{PROD}$}: BDD is more productive than UAT when producing safety test cases. \newline
\textbf{$H_1$ $_{THOR}$}: BDD yields better test thoroughness than UAT. \newline
\textbf{$H_1$ $_{FAUL}$}: BDD is more effective regarding fault detection than UAT.\newline
\textbf{$H_1$ $_{COME}$}: BDD is more effective regarding communication than UAT.\newline

\vspace{-0.2cm}
\subsection{Variables}
\vspace{-0.2cm}
The independent variables are the acceptance testing techniques. The dependent variables are: (1) productivity (PROD). It is defined as output per unit effort \cite{sdf}. In our experiment, the participants test the user stories in the STPA safety report and produce safety test cases. We assess it via the number of implemented (tested) user stories\footnote{\scriptsize{In this article, user stories are safety-related user stories.}} per minute (NIUS) \cite{sdf}; (2) test thoroughness (THOR). Code coverage is an important measure for the thoroughness of test suites including safety test suites \cite{marick1999misuse}. Considering a low complexity of our provided systems, line coverage (LC) \cite{madeyski2010impact} is more suitable than branch coverage (BC); (3) fault detection effectiveness (FAUL). Mutation testing \cite{hamlet1977testing} is powerful and effective to indicate the capability at finding faults \cite{madeyski2010impact}. In our experiment, we measure how well a safety test suite is able to find faults at the code level. We assess this via a Mutation Score Indicator (MSI) \cite{madeyski2010impact}; (4) communication effectiveness (COME). We assess this via a post-questionnaire with 11 questions for developers covering topics like understandability and 13 questions for business analysts covering topics like confidentiality according to Adzic \cite{adzic2009bridging}. The results are in a 5-point scale from -2 (negative) to +2 (positive).

\vspace{-0.2cm}
\subsection{Pilot study}
\vspace{-0.2cm}
Six master students majoring in software engineering took part in a pilot study. We arranged a four-hour training program. The first author observed the operation and concluded as follows: (1) The STPA safety report was too complicated to be used by inexperienced students. We used a comprehensive STPA report by using XSTAMPP\footnote{\scriptsize{http://www.xstampp.de/}} in the pilot study. However, a lot of unnecessary data, such as accidents, hazards and safety requirements at the system level, influenced the understanding. It costs too much time to capture the information. Thus, we simplified the STPA report with the process variables, algorithms, and UCAs. (2) We used the original Java code from a previous student project. The complex code affected the quick understanding. After the pilot study, we simplified it. (3) Training is extremely important. In the pilot study, one participant had not taken part in the training program, which led to his experiment being unfinished. We provide a textual tutorial and system description for each participant as a backup. (4) We have only used an experiment report to record the measures. However, the pure numbers sometimes cannot show clear causalities. Thus, we use a screen video recording in parallel with the experiment report.  \newline

\vspace{-0.24cm}
\subsection{Experiment operation}
\vspace{-0.2cm}
\begin{figure*}[!h] 
\centering
\small
\includegraphics[width=1.0\textwidth]{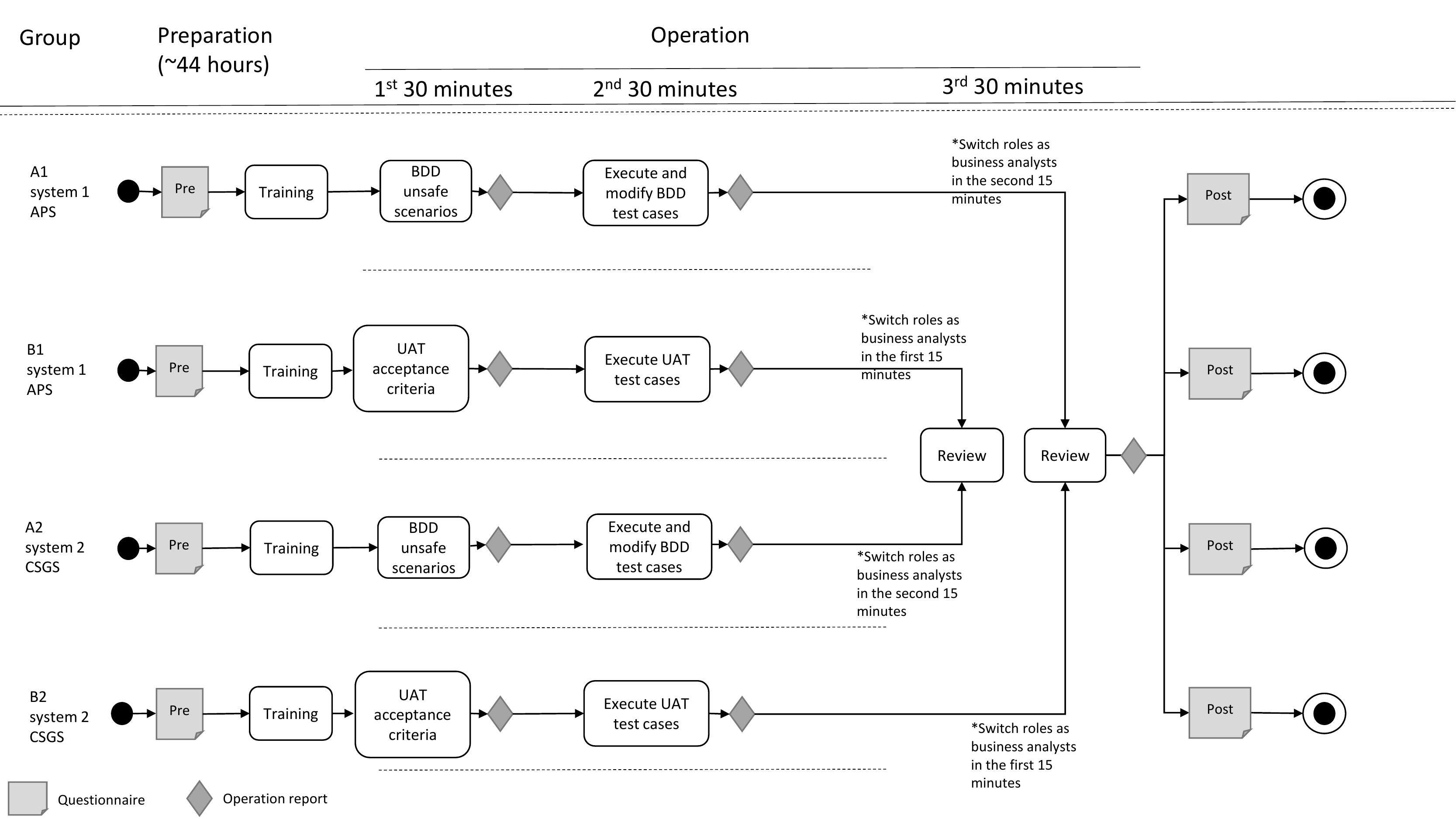} 
\caption{Experiment operation} 
\end{figure*}

As we can see in Fig. 3, we divide the 44 participants into 4 groups. We provide 2 systems and evaluate 2 acceptance testing methods. Group A1 uses BDD for system 1. Group A2 uses BDD for system 2. Group B1 uses UAT for system 1. Group B2 uses UAT for system 2. We use two systems to evaluate the communication between developers and business analysts. The developers are the participants in each group, while the fictional business analysts are portrayed by the participants in the other group using various testing methods and systems. \par   
The experiment consists of 2 phases: \emph{preparation} and \emph{operation}. The \emph{preparation} was run 2 weeks before the experiment to perform the pre-questionnaire and training.  
The \emph{operation} consists of three sessions (30 minutes/session).  
In the $1^{st}$ session, four groups write acceptance test cases. Group A1 (BDD) and Group A2 (BDD) write test scenarios in Eclipse with the Jbehave plug-in as a story file. Group B1 (UAT) and Group B2 (UAT) write acceptance criteria in plaintext. We provide 30 unsafe control actions (UCAs) in an STPA safety report. When the students finish all the 30 UCAs in 30 minutes, they record the time in minutes. After the $1^{st}$ session, the participants record the NIUS and the time in the operation report. 
In the $2^{nd}$ session, Group A1 (BDD) and Group A2 (BDD) write each test scenario into a test case and run the test case. If it fails, they should modify the trigger (code) and pass the test case. Group B1 (UAT) and Group B2 (UAT) review Java code, execute the test cases manually and complete their acceptance test report. At the end of the $2^{nd}$ session, they run PIT mutation testing. The LC and MSI are generated automatically in the PIT test report. They write down the results in the operation report. 
In the $3^{rd}$ session, the participant portrays as a developer for 15 minutes and a business analyst for 15 minutes. The developer is expected to explain his/her testing strategy as clearly as possible, while the fictional business analyst should try to question the developer. To this end, they answer a post-questionnaire.

\vspace{-0.2cm}

\begin{table}[!htb]
\centering
\begin{threeparttable}
\caption{Descriptive Statistic}
\scriptsize
\begin{tabular}{ccccclclcc}
\toprule
  Measure & Treatment & Experiment & Mean & St.Dev & Min & Median & Max & 95\% CI lower & 95\% CI upper  \\ \hline 
  NIUS    & BDD & Group A1 & 0.52 & 0.24 &  0.26 & 0.45 & 1.20 & 0.37 & 0.66 \\ 
          &     & Group A2 & 0.69 & 0.19 &  0.42 & 0.65 & 1.00 & 0.58 & 0.80 \\ 
          & UAT & Group B1 & 0.58 & 0.22 &  0.33 & 0.57 & 1.00 & 0.45 & 0.71 \\ 
          &     & Group B2 & 0.67 & 0.29 &  0.27 & 0.60 & 1.20 & 0.50 & 0.84 \\ \hline
 LC      & BDD & Group A1 & 0.02 & 0.01 &   0.01 & 0.02 & 0.05 & 0.02 & 0.03 \\ 
          &     & Group A2 & 0.02 & 0.01 &  0.01 & 0.02 & 0.04 & 0.02 & 0.03 \\ 
          & UAT & Group B1 & 0.02 & 0.01 &  0.01 & 0.01 & 0.03 & 0.01 & 0.02 \\ 
          &     & Group B2 & 0.02 & 0.01 &  0.01 & 0.01 & 0.03 & 0.01 & 0.02 \\ \hline
 MSI      & BDD & Group A1 & 0.90 & 0.38 &  0.36 & 1.00 & 1.33 & 0.67 & 1.13 \\ 
          &     & Group A2 & 0.93 & 0.49 &  0.44 & 0.83 & 2.17 & 0.63 & 1.22 \\ 
          & UAT & Group B1 & 0.89 & 0.36 &  0.42 & 0.88 & 1.56 & 0.67 & 1.10 \\ 
          &     & Group B2 & 0.85 & 0.46 &  0.30 & 0.65 & 1.63 & 0.58 & 1.12 \\ \hline
 COME     & BDD & Group A1 & 1.27 & 0.81 &  -2.00 & \textbf{1.50} & 2.00 & 0.79 & 1.75 \\ 
          &     & Group A2 & 1.18 & 0.70 &  -1.00 & \textbf{1.00} & 2.00 & 0.76 & 1.58 \\ 
          & UAT & Group B1 & -0.05 & 1.20 & -2.00 & \textbf{0.00} & 2.00 & -0.75 & 0.66 \\ 
          &     & Group B2 & 0.01 & 1.13 &  -2.00 & \textbf{0.50} & 2.00 & -0.67 & 0.67 \\ 
 \toprule
\end{tabular}
\begin{tablenotes}
      \tiny
      \item Note: St. Dev means standard deviation; CI means confidence interval. NIUS means number of implemented (tested) user stories per minute. LC means line coverage. MSI means mutation score indicator. COME was assessed via questionnaire with the results in a 5-point scale from -2 (negative) to +2 (positive).   
    \end{tablenotes}
\end{threeparttable}
\end{table}
\vspace{-0.2cm}
\vspace{-0.2cm}
\section{Analysis}
\vspace{-0.2cm}

\subsection{Descriptive analysis}
\vspace{-0.2cm}
\begin{figure*}[!htb] 
\centering
\small
\subfloat[NIUS]{%
\includegraphics[width=.30\textwidth]{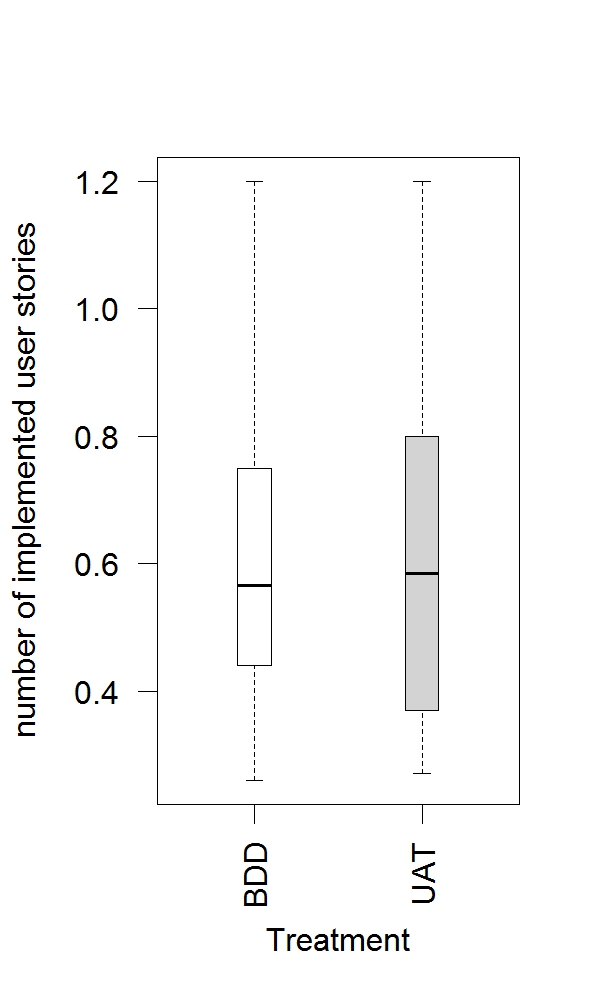}}
\subfloat[LC]{%
\includegraphics[width=.30\textwidth]{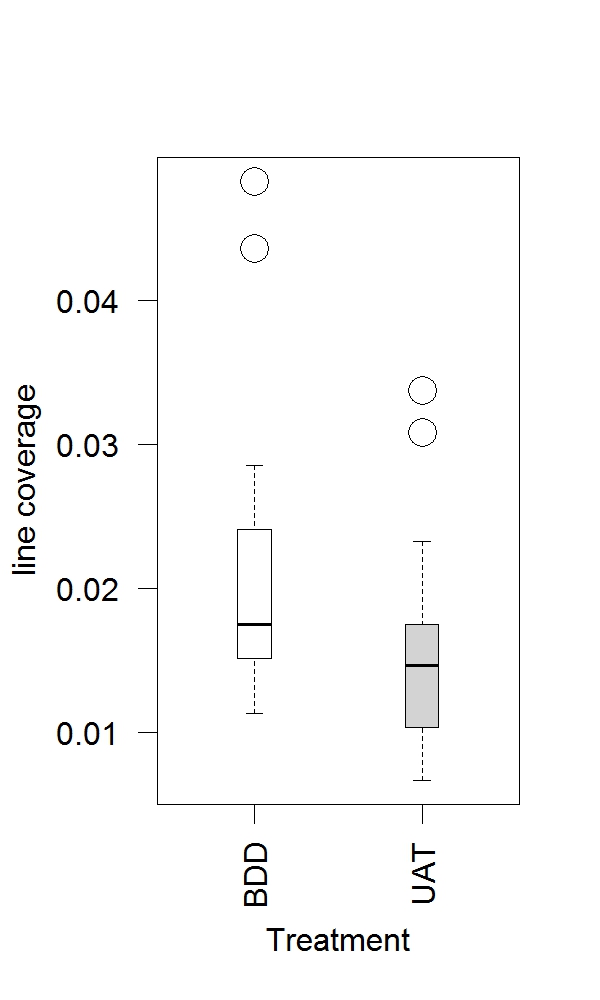}}
\subfloat[MSI]{%
\includegraphics[width=.30\textwidth]{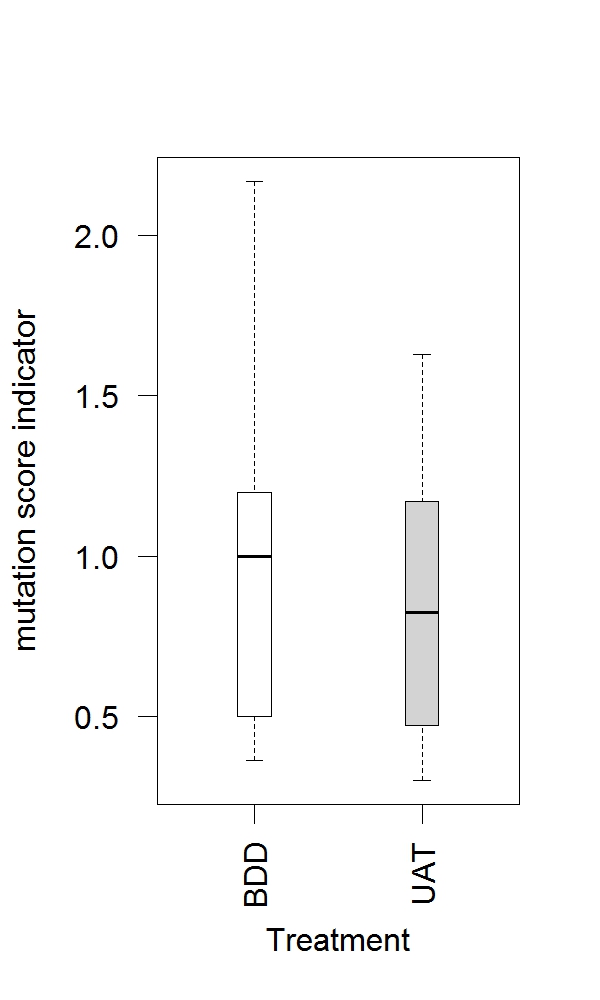}}\\
\caption{Boxplot for PROD, THOR and FAUL} 
\end{figure*}

In Table 1, we summarize the descriptive statistics of the gathered measures\footnote{\scriptsize{Raw data is available online: https://doi.org/10.5281/zenodo.1154350}}. To sum up, the results from the two systems in one treatment are almost identical. BDD and UAT have only small differences regarding NIUS and MSI. However, COME in BDD (Mean = 1.27, 1.18; Std.Dev = 0.81, 0.70) and UAT (Mean = -0.05, 0.01; Std.Dev = 1.20, 1.13) differ more strongly. LC has a small difference. In Fig. 4, we show a clear comparison and can see some outliers concerning LC. In Fig. 5, we use an alluvial diagram to show COME. We can conclude that BDD has a better communication effectiveness than UAT from the perspective of developers and business analysts respectively (depending on the length of black vertical bar on the right side of Fig. 5 (a) and Fig. 5 (b)). On the left side, we list 24 sub-aspects of assessing the communication effectiveness. The boldness of the colorful lines indicates the degree of impact. A thicker line has a larger impact on each aspect. We can see six noteworthy values from Fig. 5 (a) that BDD is better than UAT: (4) Test cases have a clear documentation. (5) They could flush out the functional gaps before development. (6) They have a good understanding of business requirements. (7) Test cases have a good organization and structure. (8) Realistic examples make them think harder. (11) There is an obvious glue between test cases and code. From Fig. 5 (b), five noteworthy values show that BDD is better than UAT: (6) The developers consider safety requirements deeply and initially. (8) It is easy to identify conflicts in business rules and test cases. (9) They are confident about the test cases. (12) They are clear about the status of acceptance testing. (13) They could spend less time on sprint-end acceptance testing but more in parallel with development. In addition, the other aspects show also slightly better results when using BDD than UAT. 

\vspace{-0.2cm}
\begin{figure*}[!htb] 
\centering

\subfloat[Developer's perspective]{%
\includegraphics[width=.48\textwidth]{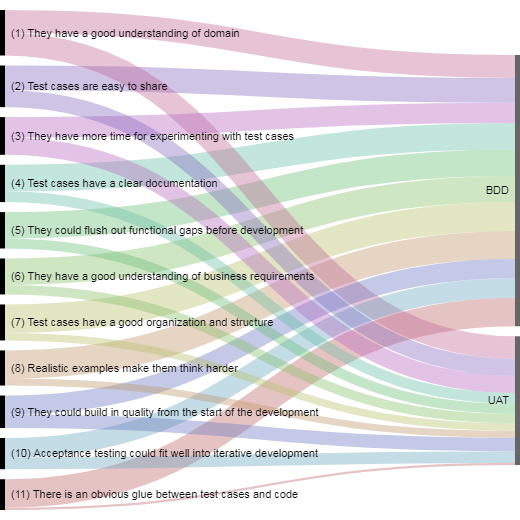}}\hfill
\subfloat[Business analyst's perspective]{%
\includegraphics[width=.48\textwidth]{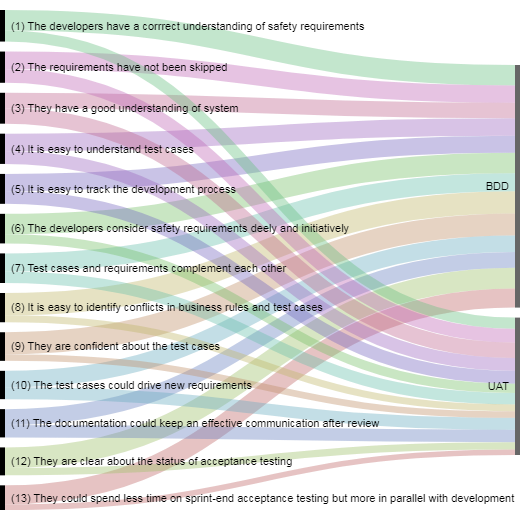}}\\
\caption{Alluvial diagram for communication effectiveness} 
\end{figure*}

\vspace{-0.2cm}
\subsection{Hypothesis testing}
\vspace{-0.2cm}
To start with, we evaluate the pre-questionnaire. No statistically significant differences between BDD and UAT groups are found concerning Java programming, acceptance testing, knowledge on SCS and agile techniques (t-test, $\alpha$ = 0.05, p $>$ 0.05 for all test parameters). 
Furthermore, we test the normality of the data distribution with Kolmogorov-Smirnov and Shapiro-Wilk tests at $\alpha$ = 0.05. The results show that the data for NIUS in Group A1, for LC in Group A1, A2, B2 and for MSI in Group A1, A2 are not normally distributed. Thus, we use non-parametric tests in the analysis.  
In addition to the use of p-values for hypotheses testing ($\alpha$ = 0.05, one-tailed) from the Mann-Whitney test, Wilcoxon test and ANOVA test, we include the effect size Cohen's d. Since we expect BDD to be better than UAT, we use one-tailed tests. NIUS is not significantly affected by using the BDD or the UAT approach (system 1: p=0.206; system 2: p=0.359, non-significant). LC is not significantly affected by using BDD or UAT (system 1: p=0.057; system 2: p=0.051, non-significant). MSI shows no statistically significant difference between using BDD or UAT (system 1: p=0.472; system 2: p=0.359, non-significant). However, COME is significantly different (system 1: p$<$0.00001; system 2: p$<$0.00001, significant). We accept the alternative hypothesis that BDD shows better communication effectiveness than UAT.    
Cohen's d shows the values around 0.2, which signifies small effects, around 0.5 stands for medium effects and around 0.8 for large effects. Thus, for COME, system 1 shows a large effect (d = 2.908). For LC we have both medium effects (system 1: d = 0.684; system 2: d = 0.662). The rest of the effects are small.

\vspace{-0.2cm}
\section{Threats to validity}
\vspace{-0.2cm}

\subsection{Internal validity}
\vspace{-0.2cm}
\emph{First}, note that we have four groups in our experiment. To avoid a multiple group threat, we prepare a pre-questionnaire to investigate the students' background knowledge. The results of the t-tests show no statistically significant differences among the groups concerning each measure. \emph{Second}, concerning the instrument, UAT is faster to learn than BDD regarding the use of tools. Even though we provide a training to narrow the gap, the \emph{productivity} might have been influenced, since the students have to get familiar with the hierarchy of writing test suites in a BDD tool. The artifacts, such as tutorials and operation report, are designed respectively with the same structure to avoid threats. In addition to the observation, we save the participants' workspaces after the experiment and video recordings for deep analysis. \emph{Third}, the students majoring in software engineering might identify more with the developer role than the business analyst role. Thus, we design two comparable systems. The students in each pair use different systems and test approaches to reduce the influence of prior knowledge. Moreover, we provide a reference \cite{gregorio2012business} on how to perform as a business analyst in an agile project. We also mention their responsibilities in the training.  
\vspace{-0.2cm}
\subsection{Construct validity}
\vspace{-0.2cm}
\emph{First}, the execution of BDD is a variant. BDD should begin with writing tests before coding. However, in our experiment, we use BDD for test-last acceptance testing rather than test-driven design. Thus, we provide source code with mutants. The measures we used could be influenced. In BDD test-first, we write failing test cases first and work on passing all of them to drive coding. According to \cite{huang2009empirical} \cite{rafique2013effects}, BDD test-first might be as effective as or even more effective than BDD test-last. \emph{Second}, the evaluation concerning productivity, test thoroughness, fault detection effectiveness and communication effectiveness does not seem to be enough. As far as we know, our study is the first controlled experiment on BDD. We can base our measurement (PROD, THOR, FAUL) mainly on TDD controlled experiments and some limited experiments on safety verification. There might be better ways to capture how well safety is handled in testing.
\vspace{-0.2cm}
\subsection{Conclusion validity}
\vspace{-0.2cm}
\emph{First}, concerning violated assumptions of statistical tests, the Mann-Whitney U-test is robust when the sample size is approximately 20. For each treatment, we have 22 students. Moreover, we use Wilcoxon W test as well as Z to increase the robustness. Nevertheless, under certain conditions, non-parametric rank-based tests can themselves lack robustness \cite{kitchenham2017}. \emph{Second}, concerning random heterogeneity of subjects, we arranged them based on the Java programming experience. According to the pre-questionnaire, the students are from the same course and 88.6\% of them are in the same major. 
\vspace{-0.2cm}
\subsection{External validity}
\vspace{-0.2cm}
\emph{First}, the subjects are students. Although there are some skilled students who could perform as well as experts, most of them lack professional experience. This consideration may limit the generalization of the results. To consider this debatable issue in terms of using students as subjects, we refer to \cite{falessi2017empirical}. They said: conducting experiments with professionals as a first step should not be encouraged unless high sample sizes are guaranteed. In addition, a long learning cycle and a new technology are two hesitations for using professionals. STPA was developed in 2012, so there is still a lack of experts on the industrial level. BDD has not been used for verifying safety requirements. Thus, we believe that using students as subjects is a suitable way to aggregate contributions in our research area. We also refer to a study by Cleland-Huang and Rahimi, which successfully ran an SCS project with graduate students \cite{cleland2017case}. \emph{Second}, the simplicity of the tasks poses a threat. We expect to keep the difficulty of the tasks in accordance with the capability of students. Nevertheless, the settings are not fully representative of a real-world project.            

\vspace{-0.2cm}
\section{Discussion \& Conclusion}
\vspace{-0.2cm}

The main benefit of our research is that we propose a possible way to use BDD for safety verification with STPA for safety analysis in agile development. We validate the combination in a controlled experiment with the limitation that we used BDD only in a test-last way. The experiment shows some remarkable results. The \emph{productivity} has no statistically significant difference between BDD and UAT. That contradicts our original expectation. We would expect BDD, as an automated testing method, to be more productive than manual UAT. Yet, as the students are not experts in our experiment, they need considerable time to get familiar with the BDD tool. The students use Jbehave to write BDD test cases in our experiment, which has strict constraints on hierarchy and naming conventions to connect test scenarios with test cases. UAT should be easier to learn. We therefore analyzed our video recordings and found that BDD developers use nearly 25\% to 50\% of their time to construct the hierarchy and naming. Scanniello et al. \cite{scanniello2016students} also mentioned this difficulty when students apply TDD. In the future, we plan to use skilled professionals in test automation to replicate this study. This could lead to different results. The \emph{test thoroughness} and \emph{fault detection effectiveness} show a non-significant difference between BDD and UAT. We could imagine that our provided Java code is too simplified to show a significant difference. The mutants are easily found with a review. These aspects need further research. \par 
The \emph{communication effectiveness} shows better results by using BDD than UAT on 24 aspects. We highlight 11 significant aspects. 
The \emph{developers} found that: \textbf{BDD has a clear documentation.} A clear documentation of acceptance test cases is important for communication \cite{borge2012acceptance}. The scenarios are written in plain English with no hidden test instrumentation. The given-when-then format is clear for describing test scenarios for safety verification based on system theory. 
\textbf{The developers using BDD could flush out functional gaps before development.} The communication concerning safety could happen at the beginning of the development. They discuss safety requirements with the business analysts and spot the detailed challenges or edge cases before functional development. UAT happens mostly at the end of the development. It makes the rework expensive and is easy to be cut in safety-critical systems.    
\textbf{The developers using BDD have a good understanding of the business requirements.} A good understanding of safety requirements helps an effective communication. They could build a shared understanding in the ``3 Amigos Meeting" to ensure that their ideas about the safety requirements are consistent with the business analysts. The developers using UAT might understand safety requirements with a possible bias. 
\textbf{BDD test cases have a good organization and structure.} This makes the test cases easy to understand, especially during maintenance. They include strict naming conventions and a clear hierarchy to manage test scenarios and test cases. 
\textbf{Realistic examples in BDD make the developers think harder.} The safety requirements are abstract with possibly cognitive diversity, which leave a lot of space for ambiguity and misunderstanding. That negatively influences effective communication. Realistic examples give us a much better way to explain how safe scenarios really work than pure safety requirements do.  
\textbf{There is an obvious glue between BDD test cases and code.} There is glue code in BDD safety verification, which allows an effective separation between safety requirements and implementation details. This glue code supports the understanding and even communication between business analysts and developers. In addition, it ensures the bidirectional traceability between safety requirements and test cases. 
The \emph{business analysts} thought that: 
\textbf{The developers using BDD consider the safety requirements deeply and initiatively.} The collaboration promotes a sense of ownership of the deliverable products. That increases an initiative communication. Instead of passively reading the documents, the developers participate in the discussion about writing test scenarios and are more committed to them.   
\textbf{The business analysts are more confident about the BDD test cases.} Confidence promotes effective communication \cite{adler1977conficence}. The business analysts could give a big picture with safety goals to the developers. Feedback from developers and their realistic unsafe scenarios give the business analysts confidence that the developers understand the safety goals correctly. 
\textbf{It is easy to identify conflicts in business rules and test cases when using BDD.} BDD has a set of readable test scenarios focusing on business rules (safety requirements). Each test scenario and test case are directly connected to the code. The business analysts can pull out test cases related to a particular business rule. This helps communication, especially when there is a changing request.  
\textbf{The business analysts are clear about the status of acceptance testing when using BDD.} It promotes a state-of-art communication. That can be attributed to the automated test suites, which might be connected with a continuous integration server and a project management tool to receive a verification report automatically.   
\textbf{The business analysts could spend less time on sprint-end acceptance tests but more in parallel with development.} They can verify the safety requirements periodically and therefore enhance communication throughout the project. \par
In conclusion, to some extent, BDD is an effective method for verifying safety requirements in agile development. As this is the first experiment investigating BDD for safety verification, further empirical research is needed to check our results. We invite replications of this experiment using our replication package\footnote{\scriptsize{https://doi.org/10.5281/zenodo.846976}}.

%
%


\begin{thebibliography}{6}

%
\bibitem{dybaa2008empirical}Dyb{\aa}, T. and Dings{\o}yr, T.: Empirical studies of agile software development: A systematic review. Information and Software Technology 50(9-10) (2008): 833-859.

\bibitem{cleland2017case}Cleland-Huang, J. and Rahimi, M.: A case study: Injecting safety-critical thinking into graduate software engineering projects. Proceedings of the 39th International Conference on Software Engineering: Software Engineering and Education Track. IEEE, 2017.

\bibitem{arthur2017applying}Arthur, J.D. and Dabney. J.B.: Applying standard independent verification and validation (IV\&V) techniques within an Agile framework: Is there a compatibility issue?. Proceedings of Systems Conference, IEEE, 2017.

\bibitem{fleming2015safety}Fleming, C.: Safety-driven early concept analysis and development. Dissertation. Massachusetts Institute of Technology, 2015.

\bibitem{wang2016toward}Wang, Y. and Wagner, S.: Toward integrating a system theoretic safety analysis in an agile development process. Proceedings of Software Engineering, Workshop on Continuous Software Engineering, 2016.

\bibitem{leveson2011engineering}Leveson, N.: Engineering a safer world: Systems thinking applied to safety. MIT Press, 2011.

\bibitem{wang2016towards}Wang, Y. and Wagner, S.: Towards applying a safety analysis and verification method based on STPA to agile software development. IEEE/ACM International Workshop on Continuous Software Evolution and Delivery, IEEE, 2016.

\bibitem{martins2017requirements}Martins, L.E. and Gorschek, T.: Requirements engineering for safety-critical systems: Overview and challenges. IEEE Software, 2017.

\bibitem{vuori2011agile}Vuori, M.: Agile development of safety-critical software. Tampere University of Technology. Department of Software Systems, 2011.

\bibitem{staalhane2012application}St{\aa}lhane, T., Myklebust, T. and Hanssen, G.K.: The application of Safe Scrum to IEC 61508 certifiable software. Proceedings of the 11th International Probabilistic Safety Assessment and Management Conference and the Annual European Safety and Reliability Conference, 2012.

\bibitem{ge2010iterative}Ge, X., Paige, R.F. and McDermid, J.A.: An iterative approach for development of safety-critical software and safety arguments. Proceedings of Agile Conference. IEEE, 2010.


\bibitem{wang2017exploratory}Wang, Y., Ramadani, J. and Wagner, S.: An exploratory study of applying a Scrum development process for safety-critical systems. Proceedings of the 18th International Conference on Product-Focused Software Process Improvement, 2017.

\bibitem{eleftherakis2003agile}Eleftherakis, G. and Cowling, A.J.: An agile formal development methodology. Proceedings of the 1st South-East European Workshop on Formal Methods. 2003.

\bibitem{ghezzi2013requirements}Ghezzi, C. et al.: On requirements verification for model refinements. Proceedings of Requirements Engineering Conference. IEEE, 2013.

\bibitem{wynne2012cucumber}Wynne, M. and Hellesoy, A.: The cucumber book: Behaviour-driven development for testers and developers. Pragmatic Bookshelf, 2012.

\bibitem{smart2015bdd}Smart, J. F.: BDD in action: Behavior-driven development for the whole software lifecycle. Manning, 2015. 

\bibitem{silva2017behavior}Silva, T.R., Hak, J.L. and Winckler, M.: A behavior-based ontology for supporting automated assessment of interactive systems. Proceedings of the 11th International Conference on Semantic Computing. IEEE, 2017.

\bibitem{hummel2013role}Hummel, M., Rosenkranz, C. and Holten, R.: The role of communication in agile systems development. Business \& Information Systems Engineering 5.5 (2013): 343-355.

\bibitem{okubo2014security}Okubo, T. et al.: Security and privacy behavior definition for behavior driven development. Proceedings of the 15th International Conference on Product-Focused Software Process Improvement. 2014.

\bibitem{lai2014combining}Lai, S.T., Leu, F.Y. and Chu, W.: Combining IID with BDD to enhance the critical quality of security functional requirements. Proceedings of the 9th International Conference on Broadband and Wireless Computing, Communication and Applications. IEEE, 2014.

\bibitem{fucci2013replicated}Fucci, D. and Turhan, B.: A replicated experiment on the effectiveness of test-first development. Proceedings of the International Symposium on Empirical Software Engineering and Measurement. IEEE, 2013.

\bibitem{fucci2016dissection}Fucci, D. et al.: A dissection of test-driven development: Does it really matter to test-first or to test-last?. IEEE Transactions on Software Engineering 43.7 (2017): 597-614.

\bibitem{sdf}Erdogmus, H., Morisio, M. and Torchiano, M.: On the effectiveness of the test-first approach to programming. IEEE Transactions on Software Engineering 31.3 (2005): 226-237.

\bibitem{kollanus2008understanding}Kollanus, S. and Isom{\"o}tt{\"o}nen, V.: Understanding \protect{TDD} in academic environment: experiences from two experiments. Proceedings of the 8th International Conference on Computing Education Research. ACM, 2008.

\bibitem{hamlet1977testing}Hamlet, R.G.: Testing programs with the aid of a compiler. IEEE Transactions on Software Engineering 4 (1977): 279-290.

\bibitem{madeyski2010impact}Madeyski, L.: The impact of test-first programming on branch coverage and mutation score indicator of unit tests: An experiment. Information and Software Technology 52.2 (2010): 169-184.

\bibitem{marick1999misuse}Marick, B.: How to misuse code coverage. Proceedings of the 16th International Conference on Testing Computer Software, 1999.

\bibitem{panvcur2011impact}Pan{\v{c}}ur, M. and Ciglari{\v{c}}, M.: Impact of test-driven development on productivity, code and tests: A controlled experiment. Information and Software Technology 53.6 (2011): 557-573.

\bibitem{george2004structured}George, B. and Williams, L.: A structured experiment of test-driven development. Information and Software Technology 46.5 (2004): 337-342.

\bibitem{siniaalto2007comparative}Siniaalto, M. and Abrahamsson, P.: A comparative case study on the impact of test-driven development on program design and test coverage. Proceedings of 1st International Symposium on Empirical Software Engineering and Measurement. 2007.


\bibitem{north2012jbehave}North, D.: JBehave. A framework for behaviour driven development. 2012.

\bibitem{wohlin2012experimentation}Wohlin, C. et al.: Experimentation in software engineering. Springer Science \& Business Media, 2012.

\bibitem{falessi2017empirical}Falessi, D. et al.: Empirical software engineering experts on the use of students and professionals in experiments. Empirical Software Engineering 23.1 (2018): 452-489.

\bibitem{enoiu2016controlled}Enoiu, E.P. et al.: A controlled experiment in testing of safety-critical embedded software. Proceedings of the International Conference on Software Testing, Verification and Validation. IEEE, 2016.

\bibitem{adzic2009bridging}Adzic, G.: Bridging the communication gap: specification by example and agile acceptance testing. Neuri Limited, 2009.

\bibitem{gregorio2012business}Gregorio, D.: How the business analyst supports and encourages collaboration on agile projects. Proceedings of International Systems Conference. IEEE, 2012.

\bibitem{scanniello2016students}Scanniello, G. et al.: Students' and professionals' perceptions of test-driven development: a focus group study. Proceedings of the 31st Annual Symposium on Applied Computing. ACM, 2016.

\bibitem{crispin2009agile}Crispin, L. and Gregory, J.: Agile testing: A practical guide for testers and agile teams. Pearson Education, 2009.

\bibitem{huang2009empirical}Huang, L. and Holcombe, M.: Empirical investigation towards the effectiveness of Test First programming. Information and Software Technology 51.1 (2009): 182-194.

\bibitem{madeyski2008impact}Madeyski, L.: Impact of pair programming on thoroughness and fault detection effectiveness of unit test suites. Software Process: Improvement and Practice 13.3 (2008): 281-295.

\bibitem{rafique2013effects}Rafique, Y. and Mi{\v{s}}i{\'c}, V.B.: The effects of test-driven development on external quality and productivity: A meta-analysis. IEEE Transactions on Software Engineering 39.6 (2013): 835-856.

\bibitem{borge2012acceptance}Haugset, B. and St{\aa}lhane, T.: Automated acceptance testing as an agile requirements engineering practice. Proceedings of the 45th Hawaii International Conference on System Science. IEEE, 2012.

\bibitem{adler1977conficence}Adler, R. B.: Confidence in communication: A guide to assertive and social skills. Harcourt School, 1977.

\bibitem{kitchenham2017}Kitchenham, B. et al.: Robust statistical methods for empirical software engineering. Empirical Software Engineering 22.2 (2017): 579-630.

\end{thebibliography}
\end{document}